\documentclass[pra,showpacs,twocolumn,floatfix]{revtex4}
\usepackage{amsmath}
\usepackage{mathrsfs}
\usepackage{graphicx}

\newcommand{\ket}[1]{|\hspace{0.5pt}#1\hspace{0.5pt}\rangle}
\newcommand{\bra}[1]{\langle\hspace{0.5pt}#1\hspace{0.5pt}|}
\newcommand{\bracket}[--2]{\langle\hspace{0.5pt}#1\hspace{0.5pt}|\hspace{0.5pt}
    #2\hspace{0.5pt}\rangle}
\renewcommand{\v}[1]{\ensuremath{\mathbf{#1}}}
\begin{document}
\title{Topological Entropy of Quantum Hall States in  Rotating Bose Gases}
\author{Alexis G. Morris}
\author{David L. Feder}
\affiliation{Department of Physics and Astronomy and Institute for Quantum
Information Science, University of Calgary, Calgary, Alberta, Canada T2N 1N4}
\begin{abstract}
Through exact numerical diagonalization, the von Neumann entropy is calculated
for the Laughlin and Pfaffian quantum Hall states in rotating interacting Bose 
gases at zero temperature in the lowest Landau level limit. The particles 
comprising the states are indistinguishable, so the required spatial 
bipartitioning is effected by tracing over a subset of single-particle 
orbitals. The topological entropy is then extracted through a finite-size 
scaling analysis. The results for the Laughlin and the Pfaffian states agree 
with the expected values of $\ln\sqrt{2}$ and $\ln\sqrt{4}$, respectively. 
\end{abstract}

\maketitle

Quantum Hall states are characterized by topological order, in that they can 
be described by a set of quantum numbers that are robust against local 
perturbations. Examples of such topological invariants include the ground-state 
degeneracy~\cite{Wen:1990}, the Chern number~\cite{Hatsugai:1997}, and the braiding statistics of 
the quasiparticle excitations \cite{Hormozi:2007}. Interest in the topological
underpinnings of the quantum Hall effect has surged recently due to the 
possibility of using these states for performing intrinsically fault-tolerant 
quantum computation~\cite{Sarma:2007, Kitaev:2003, Hormozi:2007}. The potential 
usefulness of certain fractional quantum Hall (FQH) states for quantum 
computation stems from the expectation that their quasiparticle excitations 
possess non-Abelian fractional statistics, which can be braided in order to 
perform topologically-protected logical operations. Of the few FQH states that 
are thought to possess non-Abelian excitations (though these alone would be 
insufficient to perform universal quantum computation), the most experimentally 
accessible is the so-called Pfaffian state that occurs at a filling factor of 
$\nu_\text{F}=5/2$ in electronic FQH systems. For small particle numbers, 
calculation of the overlap between the Pfaffian wavefunction and the exact 
ground state is very 
good~\cite{Morf:1998,Rezayi:2000,Scarola:2002,Feiguin:2008}, though some 
doubt has been recently cast over the validity of the Pfaffian wavefunction 
description~\cite{Toke:2006, Toke:2007}.

FQH states have also been predicted to occur in rotating neutral Bose gases 
confined in harmonic traps
due to the formal equivalence between the Hamiltonians describing these 
systems and two-dimensional (2D) electron gases in transverse magnetic 
fields~\cite{Cooper:1999,Viefers:2000,Wilkin:2000,Regnault:2003}. Notably, the 
bosonic Laughlin state occurs at a filling factor $\nu_B=1/2$ while for the 
bosonic Pfaffian the filling factor is $\nu_B=1$.  Here the filling factor is 
defined as the ratio between number of particles and vortices.  Although such 
states have not yet been experimentally observed due to the difficulty in 
achieving the required high rotation rates~\cite{Schweikhard:2004}, bosonic
FQH states have a distinct advantage for topological quantum computing in that 
quasiparticle excitations could potentially be simple to excite and 
control~\cite{Paredes:2001}. Previous exact diagonalization calculations yield
a strong overlap between the bosonic Pfaffian wavefunction and the exact ground
state, though this decreases with increasing particle 
number~\cite{Chang:2005,Regnault:2006}. It is therefore important to calculate 
global properties such as topological quantum numbers in order to provide 
further evidence for the Pfaffian description of the ground state. 

In this work, we focus on one such quantity called the topological 
entanglement entropy $\gamma$~\cite{Levin:2006,Kitaev:2006,Li:2008}. Statistical
mechanics defines the classical entropy as proportional to the logarithm of a 
state's multiplicity. In a similar manner, $\gamma$ is defined as 
$\gamma = \ln D$, where $D\ge 1$ is the total quantum dimension of 
the topological phase~\cite{Fendley:2007}. A general quantum Hall state in the Laughlin sequence with
$\nu = 1/p$ has $D=\sqrt{p}$, while the bosonic Pfaffian state has 
$D=\sqrt{4}$. (Of course, the $\nu=1/4$ Laughlin state is
unambiguously distinguished from the Pfaffian state by the filling factor). The 
topological entropy thus provides a powerful tool for classifying different 
quantum Hall states, as long as it can readily be calculated.

Fortunately, a connection between the von Neumann entropy $S$ and topological 
order has recently been demonstrated, from which $\gamma$ can be extracted in 
principle. Suppose a topologically ordered state is separated into two 
partitions $A$ and $B$ by a circle of radius $R$. The von Neumann entropy is 
defined as $S=S_A=-{\rm Tr}\left(\rho_A\ln\rho_A\right)$, where $\rho_A$ is 
the reduced density matrix obtained after tracing over region B. In large systems, under 
conditions in which $B\gg A$ or vice versa, the von Neumann entropy will scale 
with the length $2\pi R$ of the boundary delimiting both sections as
\begin{equation}
   S=\alpha(2\pi R) - \gamma + \mathcal{O}(1/R),
   \label{scaling}
\end{equation}
where $\alpha$ is a nonuniversal coefficient.   Cutting the system into multiple 
subsections and judiciously combining the resulting von Neumann entropies, 
the terms proportional to the boundary length can be cancelled, leaving only the universal 
topological entropy $\gamma$~\cite{Kitaev:2006,Levin:2006}. In the present 
work, we use exact calculations for small number of particles and a
finite-size scaling analysis to obtain $\gamma$ for both the bosonic Laughlin 
and Pfaffian states.


We consider a zero-temperature gas of bosons confined in a cylindrically
symmetric harmonic trap that is rapidly rotated around the $z$-axis with a 
frequency $\tilde\Omega$. The axial 
trapping frequency $\omega_z$ is assumed to be much larger than that along the 
radial direction $\omega$, so that the gas can by considered 
quasi-2D. This corresponds to the disk geometry that has been 
used in previous studies of electronic FQH systems~\cite{Xie:1993}. Particles interact via a standard delta-function pseudopotential whose
strength is controlled by a 2D coupling constant 
$\tilde{g}=\sqrt{8\pi}\hbar\omega \ell^2 a/\ell_z$. Variables 
$\ell=\sqrt{\hbar/M\omega}$ and $\ell_z=\sqrt{\hbar/M\omega_z}$ are the 
characteristic oscillator lengths along the radial and axial directions, 
respectively, and $a$ is the three-dimensional scattering length. Expressing
all lengths in units of $\ell$, energies in terms of $\hbar\omega$, and 
frequencies in terms of $\omega$, the effective 2D Hamiltonian in 
the frame co-rotating with the atoms at frequency $\Omega$ is
\begin{equation*}
\hat{H}=\sum_i^N \left(-\frac{1}{2}\nabla_i^2 + \frac{1}{2}\rho_i^2\right)
-\Omega L +g\sum_{i<j}^N \delta(\v{r}_i-\v{r}_j),
\end{equation*}
where $g=\tilde{g}/\ell^2$ is the dimensionless coupling constant, $N$ is the 
number of bosons of mass $M$, $L=\sum_i^N m_i$ is the $z$-projection of the 
total angular momentum (which is a conserved quantity in this axisymmetric
potential), and $\v{r}=(\rho,\phi)$ is a particle's position in polar 
coordinates. For the remainder of this work, we set $g=1$.

Obtaining the topological entanglement entropy consists of three main steps: 
evaluation of the Hamiltonian, its diagonalization, and the calculation of the 
ground state's von Neumann entropy.  The Hamiltonian is expressed in a lowest 
Landau level (LLL) approximation~\cite{Morris:2006} Fock basis of the form
\begin{equation*}
\ket{{\cal N}_{0},{\cal N}_{1},\ldots,{\cal N}_{L}}=\prod_{m=0}^L\frac{(\hat
b_m^{\dagger})^{{\cal N}_m}}{\sqrt{{\cal N}_m!}}\ket{0},
\end{equation*}
where $\hat b_m^\dagger$ creates a boson with $m$ units of angular momentum, 
and the  $\mathcal{N}$'s are the occupation numbers. The Hilbert space size is 
determined by the number of unique ways to distribute $L$ units of angular 
momentum among $N$ particles.  This number grows very rapidly which limits our 
study to $N\le10$ for the Laughlin state and $N\le13$ for the Pfaffian state. 
The Hamiltonian is evaluated in this basis by writing it in a second quantized
form using the bosonic field operators 
$\hat \psi(\v{r})=\sum_m \hat b_m \Phi_m(\v{r})$ expanded in terms of
2D harmonic oscillator orbitals in the lowest Landau level 
approximation (i.e.\ zero principle quantum number),
$\Phi_m(\v{r})=\sqrt{\frac{1}{m!\pi}}\rho^me^{-\rho^2/2}e^{im\phi}$.  Using 
these field operators, the Hamiltonian reduces to
\begin{equation*}
\hat H=\sum_i\hat b^\dagger_i\hat b^{\vphantom{\dagger}}_i\epsilon_i
+\frac{g}{2}\sum_{ijkl}\hat b_{i}^\dagger\hat b_{j}^\dagger
\hat b^{\vphantom{\dagger}}_{k}\hat b^{\vphantom{\dagger}}_{l}\mathcal{D}_{ijkl}
\end{equation*}
where $\epsilon_i=\hbar\omega [N+L(1-\tilde\Omega)]$ and
\begin{equation*}
   \mathcal{D}_{ijkl}=\frac{\pi}{2^{L+1}}\frac{(i+j)!}{\sqrt{i!j!k!l!}}\delta_{i+j,k+l}.
\end{equation*}
In this Fock basis, the first term in $\hat{H}$ is equivalent to the identity 
matrix and simply represents an energy offset. The problem is then reduced to 
finding the eigenstates of the interaction matrix $\hat{H}_\text{int}$. These 
are obtained by exact diagonalization using a Lanczos algorithm.

A particular quantum Hall state is selected by specifying the total angular 
momentum according to the relationships between $L$ and particle number $N$ 
established in Ref.~\onlinecite{Wilkin:2000}.  The Laughlin state is selected 
by requiring that $L=N(N-1)$ while for the Pfaffian state the relation is 
$L=N(N-2)/2$ and $L=(N-1)^2/2$, for even and odd $N$ respectively. As an 
example, consider the 2-particle Laughlin state.  It occurs when $L=2(2-1)=2$ 
which corresponds to a Hilbert space spanned by the following two states: 
$\ket{101} = \hat b_2^\dagger \hat b_0^\dagger\ket{0}$ and 
$\ket{020} = \frac{1}{\sqrt{2}}\hat b_1^\dagger\hat b_1^\dagger\ket{0}$.
Once that $L$ is fixed, the Hamiltonian can be constructed and diagonalized. 
This produces a ground state wavefunction described in terms of the Fock basis 
states.  In our example, the (Laughlin) state vector obtained is
\begin{equation*}
   \ket{\Psi_\text{L}} = \frac{1}{\sqrt{2}}(\ket{101}-\ket{020}).
   \label{Laughlin1}
\end{equation*}
It is straightforward to verify that this corresponds exactly to the 
theoretical Laughlin wavefunction
\begin{equation*}
\Psi_{\rm L}=\prod_{i<j}^N (z_i-z_j)^2e^{-\sum_{i}^N|z_j|^2/2\ell^2},
\label{Laughlin}
\end{equation*}
after expanding in terms of the bosonic field operators
\begin{equation*}
   \Psi_\text{L}(\v{r}_1,\v{r}_2,\ldots,\v{r}_N)=\frac{1}{\sqrt{N!}}\bra{0}\hat\psi(\v{r}_1)\hat\psi(\v{r}_2)\ldots\hat\psi(\v{r}_N)\ket{\Psi_\text{L}}
\end{equation*}
and using $z=\rho e^{i\phi}$.

Once the ground state wavefunction has been obtained, the von Neumann entropy 
must be calculated. We adopt the method of Refs.~\cite{Haque:2007,Zozulya:2007, Friedman:2008} that uses orbital
partitioning (see Ref.~\cite{Iblisdir:2007} for particle partitioning). This consists of separating the state into regions delineated by
a specific number $n$ ($\equiv m_\text{max}+1$) of single particle orbitals, so
that region A would correspond to $m\leq m_{\rm max}$ and region B to 
$m>m_{\rm max}$. For the two-particle Laughlin state example considered above,
the reduced density matrix $\rho_A$ for $n=1$ (meaning only the $m=0$ orbital 
is in partition $A$) is
\begin{eqnarray*}
   \rho_A&=&\text{Tr}_B(\ket{\Psi_\text{L}}\bra{\Psi_\text{L}}) \\
   &=&\frac{1}{2}\left(\ket{1}\bra{1}\bracket{0}{0}\bracket{1}{1}+\ket{0}\bra{0}\bracket{2}{2}\bracket{0}{0}\right)\\
   &=&\frac{1}{2}\left(\ket{1}\bra{1}+\ket{0}\bra{0}\right)
\end{eqnarray*}
which has a two-fold degenerate eigenvalue of $1/2$. The von Neumann entropy
is thus $\rho_A=-2\times\frac{1}{2}\ln(\frac{1}{2})\approx 0.69$.

Once $S$ is obtained for $n=1$, the process is repeated numerous times for 
increasing $n$, and in principle Eq.~\eqref{scaling} can be used to obtain the 
topological entropy $\gamma$ from a plot of $S$ versus $R$.  What remains to 
be specified is the partition boundary size. The single-particle orbital 
density is ring-shaped and centered at the origin of the trap, with radius 
$\left\langle\rho\right\rangle_m=\int d{\bf r}\rho|\Phi_m(\rho,\phi)|^2\sim
\sqrt{m+1}$ for large $m$.
We therefore consider the boundary between the two regions $A$ and $B$ for a 
particular choice of $n$ to correspond to a circle of size 
$2\pi\left\langle\rho\right\rangle_{n-1}$. Since the factor of $2\pi$ can 
simply be combined with $\alpha$ in Eq.~\eqref{scaling}, we define the boundary size when the first $n$ orbitals are kept in partition $A$ as
\begin{equation}
R=\left\langle\rho\right\rangle_{n-1}.
\label{R}
\end{equation}

\begin{figure}[t]
\includegraphics[width=0.48\textwidth]{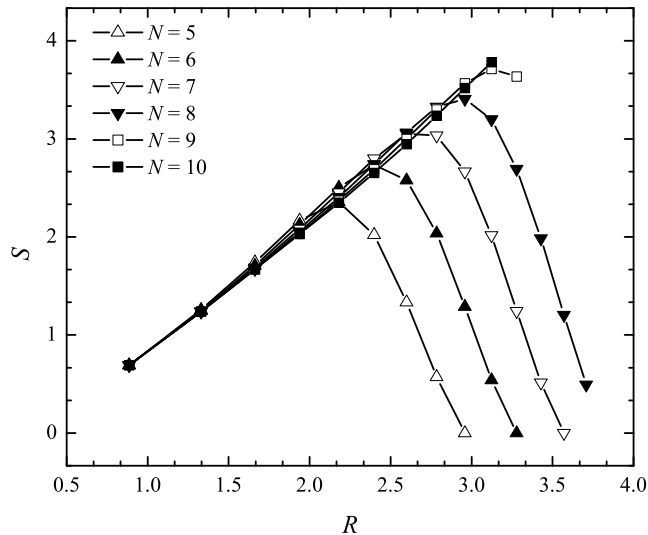}
\caption{Dependence of the von Neumann bipartite entropy $S$ with system size 
$R=\left\langle\rho\right\rangle_{n-1}$. The curves $S(R)$ are linear, as predicted by 
Eq.~\eqref{scaling}, until finite size effects begin to dominate for the 
larger partition sizes.}
\label{vonNeumann}
\end{figure}

A plot of $S(R)$ for $N=5$ to 10 is shown in Fig.~\ref{vonNeumann} for the 
bosonic Laughlin state. An initial linear increase of $S(R)$ is observed, as 
expected from Eq.~\eqref{scaling}; however, finite-size effects bring $S$ to 
zero for larger values of $R$. This is because in small-$N$ systems most of 
the particles occupy low-$m$ orbitals whose amplitudes are largest in the
vicinity of the trap center; the ground-state wavefunction has a negligeable 
overlap with higher angular momentum orbitals.

\begin{figure}[t]
\includegraphics[width=0.48\textwidth]{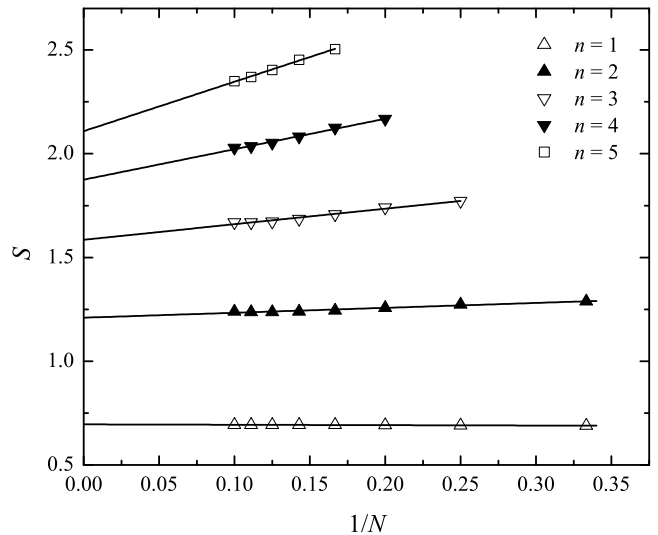}
\caption{Finite size scaling of the Laughlin state's von Neumann entropy $S$
with number of particles, for five different partition sizes. Linear 
regressions provide the $y$-intercepts which yield the values of
\mbox{$S(N\rightarrow\infty)$} shown in the inset of Fig.~\ref{topolL}.} 
\label{finitesizeL}
\end{figure}

\begin{figure}[!ht]
\includegraphics[width=0.48\textwidth]{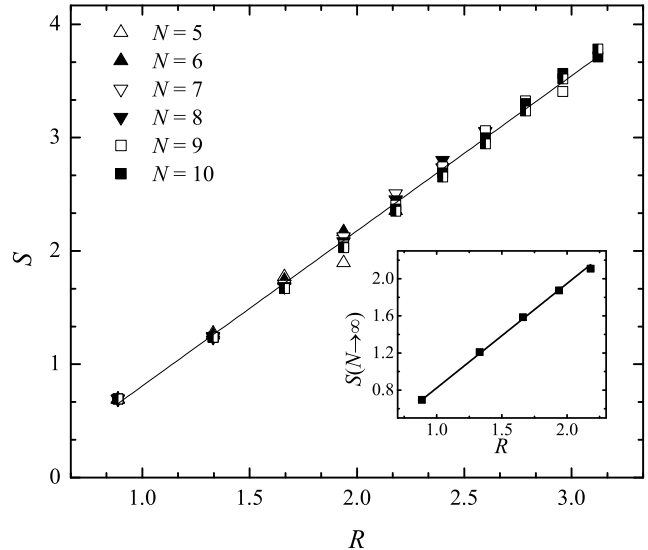}
\caption{Obtaining the Laughlin topological entropy from Eq.~\eqref{scaling} and data shown in Fig.~\ref{vonNeumann}, after removing data points clearly affected by finited size effects.  A linear fit on the combined data produces a $y$-intercept of $-0.57\pm 0.02$, leading to $\gamma_L=+0.57\pm 0.02$.   In the inset, values of $S(N\rightarrow\infty)$ obtained in Fig.~\ref{finitesizeL} are used, increasing the accuracy of the result.  A weighted linear regression yields $\gamma_L=0.30\pm 0.02$ (error bars on individual data points are smaller than the data points).}
\label{topolL}
\end{figure}

To perform a proper finite-size scaling analysis of the small-$N$ data, we 
follow the procedure introduced in Ref.~\onlinecite{Haque:2007}. The value of 
$S(N\rightarrow \infty)$ is estimated by plotting $S$ as a function of $1/N$ for region $A$ 
containing different numbers of orbitals $n$, as shown in 
Fig.~\ref{finitesizeL}. Our calculations were restricted to the range 
$1\leq n\leq 5$ because we have too little data for $n\ge 6$. To obtain the 
$N\rightarrow\infty$ values a linear regression was applied. A slight
positive curvature in the data for the largest values of $N$ might 
indicate the emergence of asymptotic values, but the trend was not clear enough 
to enable a more sophisticated analysis. The results are plotted in the inset of 
Fig.~\ref{topolL} and another linear regression is made to provide the 
$y$-intercept. We find $\gamma_L=0.30\pm  0.02$, which is slightly lower
than the expected result of 0.35 for the bosonic Laughlin state at filling $\nu=1/2$; this might reflect the rather na\"\i ve
linear analysis of the finite-size scaling.

\begin{figure}
\includegraphics[width=0.48\textwidth]{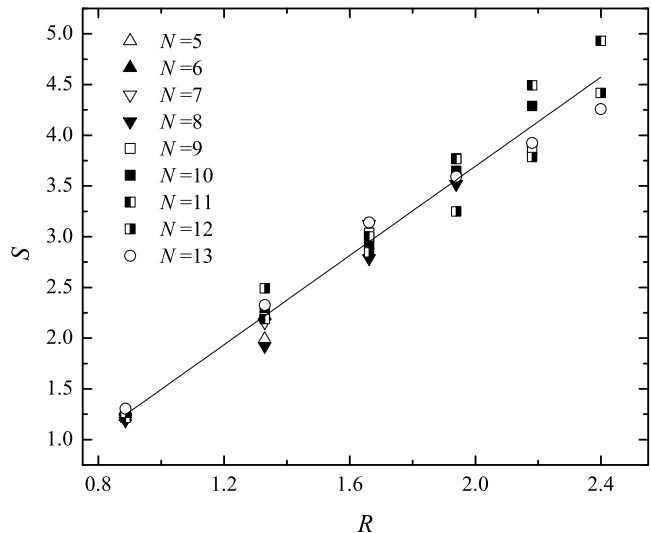}
\caption{The topological entropy for the Pfaffian state is obtained by using 
the combined data of five to thirteen particles.  Data points that were 
obviously influenced by finite size effects (for which $S(R)$ decreased with 
$R$, as in Fig.~\ref{vonNeumann}), were removed. A linear regression yields 
$\gamma_P=0.7\pm 0.1$.}
\label{topolP}
\end{figure}

The same procedure was then repeated for the Pfaffian state with $N=5$ to 13. 
Unfortunately, a larger amount of scatter was present in the Pfaffian version 
of Fig.~\ref{finitesizeL}, preventing a clean extrapolation of the $S(N)$ data 
to the large-$N$ limit. Thus, instead of plotting $S(N\rightarrow\infty)$ as a function of
$R$ to obtain $\gamma_P$, a linear regression is performed directly on the collection of data points shown in Fig.~\ref{topolP}, where $n=1$ to 6.
This approach
produces a topological entanglement entropy of $\gamma_P=0.7\pm 0.1$ 
consistent with the predicted value of $\ln(\sqrt{4})=0.69$.

The close correspondence between the Pfaffian numerical and theoretical results are quite surprinsing considering that corrections for finite-size effects have not been made.  Fitting the aggragate Laughlin results in the same manner gives
$\gamma_L=0.57\pm 0.02$ (see Fig.~\ref{topolL}) instead of the previously obtained $\gamma_L=0.30\pm 0.02$, suggesting that the obtained $\gamma_P$'s accuracy is coincidental.
The error ranges quoted
above strictly reflects the scatter in the numerical data, and underestimates
the actual uncertainty by neglecting systematic errors.  In particular, each data point is treated as equally valid,
whereas the $\mathcal{O}(1/R)$ term in Eq.~\eqref{scaling} clearly favors
large-$R$ results. Ideally, only the largest-$n$ data would have been kept;
however, this would not have provided enough data with which to extract values
of $\gamma$.  In light of the discrepancy between both Laughlin results, a more reasonable estimate of the Pfaffian topological entropy is $\gamma_P = 0.7 \pm 0.3$.


In conclusion, we have calculated the topological entanglement entropy for 
both the bosonic Laughlin and Pfaffian states for a rotating Bose gas.   For 
the Laughlin state, we obtain a result of $0.30\pm 0.02$ which is almost 
consistent with the expected value of $\ln\sqrt{2}=0.35$.  For the Pfaffian 
state we obtain $0.7\pm0.3$.  This value of the topological entropy is consistent with the expected value of $\ln\sqrt{4}=0.69$, though the large amount of scatter present in Fig.~\ref{topolP} prevents a completely unambiguous identification of the state by the Pfaffian wavefunction.

%

This work was supported by the Natural Sciences and Engineering Research 
Council of Canada, the Canada Foundation for Innovation, and Alberta's 
Informatics Circle of Research Excellence.

 \bibliographystyle{aip}
 \bibliography{/home/amorris/PhD/refs/bib}

\end{document}